# Edwards thermodynamic framework controls density segregation in cyclically sheared granular materials

Haiyang Lu,[1] Houfei Yuan,[1] Shuyang Zhang,[1] Zhikun Zeng,[1] Yi Xing,[1] Jiazhao Xu,[1] Xin Wang,[1] and Yujie Wang[1,2,3,*]

[1]*School of Physics and Astronomy, Shanghai Jiao Tong University, Shanghai 200240, China*
[2]*State Key Laboratory of Geohazard Prevention and Geoenvironment Protection, Chengdu University of Technology, Chengdu 610059, China*
[3]*Department of Physics, College of Mathematics and Physics, Chengdu University of Technology, Chengdu 610059, China*

Using X-ray tomography, we experimentally investigate granular segregation phenomena in a mixture of particles with different densities under quasi-static cyclic shear. We quantitatively characterize their height distributions at steady states by minimizing effective free energy based on a segregation temperature that captures the competition between the mixing entropy and gravitational potential energy. We find this temperature coincides with Edwards' compactivity within error under various pressures and cyclic shear amplitudes. Therefore, we find that granular segregation in quasi-static conditions can be fundamentally explained by an effective granular thermodynamic framework including real energy terms based on the Edwards statistical ensemble.

Mixtures of granular materials with varying sizes, densities, friction, inelasticity, or other properties tend to segregate under shear or shaking [1-3]. This phenomenon is ubiquitous in industrial and geophysical processes, making a fundamental understanding of it critically important. The segregation of granular mixtures can be driven by multiple mechanisms simultaneously, resulting in complex segregation behaviors due to the competition or cooperation among these mechanisms [4-6]. One primary cause of segregation is thermal effects. In the vibro-fluidized regime where granular particles are highly agitated, the kinetic temperature of the system can be defined based on the average kinetic energy of the particles [7,8]. Within such systems, a temperature gradient can arise due to nonuniform energy input or unequal energy partition among different species of particles. This temperature gradient can lead to segregation through thermal diffusion, a process well-accounted for within the framework of granular kinetic theory [9,10]. In non-fluidized systems, where particles remain in enduring contact, segregation mechanisms become more complex. Geometric segregation mechanisms, such as void-filling [11] and arching effect [12,13], have been proposed. These mechanisms are often understood phenomenologically: small particles tend to occupy voids beneath larger particles when the system is perturbed, thereby continuously elevating the larger ones. Notably, these mechanisms can also be interpreted as entropic effects [14], where small particles tend to occupy more space despite increasing the overall gravitational potential energy. It is worth noting that this entropy effect has been previously reported as the emergence of an effective depletion force in granular systems [15,16], similar to the thermal colloidal case [17]. Therefore, it is tempting to assume that a thermodynamic approach can also be developed for the enduring-contact regime, which can provide a quantitative explanation for segregation in terms of the minimization of a specific thermodynamic free energy, taking into account the competition between entropy and energy [18,19].

Back in the 1990s, Edwards and collaborators developed an effective statistical mechanics framework for jammed granular systems, where volume plays the analogous role of energy in thermodynamic systems, and compactivity $\chi$ serves as the analog of thermal temperature [20,21]. In numerical simulations on a lattice model, Nicodemi *et al.* have generalized Edwards' framework by including gravitational potential energy in the free energy to explain the segregation of hard sphere mixtures [22]. However, in this work, the segregation temperature was introduced in an *ad hoc* way. Whether the segregation can be treated within an effective thermodynamic framework in a real experimental system, and the nature of this segregation temperature, remains unknown.

In this article, we employ X-ray tomography to investigate the segregation process of two types of particles with identical sizes but different mass densities under quasistatic cyclic shear. We find that the ratio of the number distribution functions (NDFs) of two types of particles along the vertical direction follows a Boltzmann-like expression when sheared steady states are reached. By minimizing free energy, which includes the entropy of mixing and gravitational potential energy, we obtain an effective segregation temperature $T_{seg}$. We further obtain Edwards compactivity $\chi$ of the system from the microscopic volume fluctuations [23]. Our results demonstrate that $T_{seg}$ coincides with $\chi$ within experimental error, suggesting the segregation process is governed by Edwards statistics. We further propose a general free energy form for granular materials in enduring-contact regime that can incorporate real energy terms.

In the experiment, we 3D print (ProJet MJP 2500 Plus, 0.032 mm resolution) solid particles (SPs) and hollow particles (HPs) with the same plastic material (VisiJet M2R-WT, $\rho = 1.12 \times 10^3$ kg/m$^3$), where the HPs have half the mass of the SPs but identical sizes and surface properties. To prevent crystallization while ensuring a uniform particle size distribution within the system, both types of particles are designed to have diameters following a Gaussian distribution with polydispersity of 8% and $\bar{d} = 7$ mm [24]. Figure 1(a) illustrates the schematic of our experimental setup. The shear cell has a cubic shape with dimensions of 120 mm ($x$) × 120 mm ($y$) × 130 mm ($z$) at zero strain. Before each experiment, approximately 3000 SPs and 3000 HPs are uniformly mixed in the shear cell to form a packing with a height of approximately 19$d$. To investigate the influence of pressure $p$ on the segregation process, we employ two pressure conditions: under the free surface condition, the internal pressure within the packing corresponds to the hydrostatic pressure generated by the weight of particles above [25]; alternatively, under the confined condition, the packing is covered by a lid with mass $M$ of either 1.6 or 3.3 kg, approximately 2 or 4 times the total mass of the particles, ensuring nearly uniform pressure of 0.01 atm or 0.02 atm within whole packing.

The cyclic shear is generated by a step motor attached to the bottom plate of the shear cell. See Ref. [13] for detailed information on the shearing protocol. In our quasistatic cyclic shear experiment, a range of strain amplitudes $\Gamma \in [0.133, 0.33]$ are utilized. The system achieves steady-state packing fractions within several hundred shear cycles, with larger $\Gamma$ resulting in lower steady-state packing fraction [see Fig. 1(b)]. We acquire the packing structure after every 400 shear cycles when $\Gamma > 0.25$ and every 1,000 shear cycles when $\Gamma \leq 0.25$ via a medical CT scanner (UEG Medical Group Ltd., 0.2 mm spatial resolution). A total of 50 CT scans are taken in one experimental realization. Following similar image processing procedures of our previous study [13], we can distinguish between the two types of particles and simultaneously determine their individual particle centroid coordinates with an error of less than $3 \times 10^{-3} d$.

Upon shear, the initial uniform mixture of SPs and HPs tends to segregate due to their different mass densities [26,27]. To quantify the segregation process, we define the degree of segregation $\alpha = \frac{2\delta_c}{H}$, where $H$ represents the total height of the packing at steady state and $\delta_c = \overline{H_h} - \overline{H_s}$ is the difference between the average heights of HPs and SPs, respectively. According to this definition, $\alpha = 0$ indicates a uniformly mixed state, and $\alpha = 1$ represents a completely segregated state. Figures 1(c) and 1(d) show the evolution of $\alpha$ for systems under free surface and confined conditions as a function of the accumulated strain $\delta\gamma = 4\Gamma\delta n$, where $\delta n$ is the number of shear cycles. We note that all curves of $\alpha$ can be scaled by an exponential form $\alpha = 1 - \frac{\alpha - \alpha_s}{\alpha_0 - \alpha_s} = 1 - \exp(-\delta\gamma/\tau)$ [insets of Fig. 1(c) and 1(d)], regardless of the pressure conditions, where $\alpha_0$ and $\alpha_s$ are the degrees of segregation at the initial and steady states respectively, and $\tau$ corresponds to the characteristic time scale of the segregation (see Supplemental Material [28] for more details). As noted before, the time scale associated with the segregation is significantly larger than that for the system to reach steady-state packing fractions, typically achieved within a few hundred cycles [Fig. 1(b) and its inset]. This suggests that the segregation predominantly occurs after the system has already reached its steady-state packing fraction (as clarified later, corresponding to segregation occurring within Edwards "thermodynamic" equilibrium). Once the steady state of segregation is reached, the NDFs, $n_\gamma(H)$ with $\gamma \in [s, h]$ corresponding to SPs and HPs, also exhibit steady distributions. Here $n_\gamma(H)$, defined as the number density (in unit of $d^3$) at different heights $H$, is obtained by averaging the data over horizontal slices of the sample of thickness $2d$. [see Figs. 2(a) and 2(b)]. These distributions bear a striking resemblance to the height-dependent density distribution of a thermal gas under gravity, $\rho(H) = \rho_0 e^{-mgH/k_B T}$ [29], suggesting a potentially analogous thermodynamic origin. To elucidate the steady-state distributions of $n_s(H)$ and $n_h(H)$, we introduce an effective thermodynamic framework [22]. In this framework, the free energy is formulated as $F = E - T_{seg}S$, where $E$ is the gravitational potential energy, $S$ is the total entropy, and $T_{seg}$ is an undetermined segregation temperature. In principle, $T_{seg}S$ term favors the mixing of different-density particles, while gravitational potential energy tends to segregate them. In practice, when calculating the entropy, we only consider the entropy of mixing since the configurational entropy associated with the disordered structures remains constant once the global packing fraction reaches its steady-state value. This is because the configurational entropy does not change with segregation, given that both types of particles have identical sizes and surface properties. Subsequently, the entropy and the gravitational potential energy for the system can be expressed respectively as:

$$S = -2\int_0^{H_{max}} n_s(H)\ln n_s(H) dH - 2\int_0^{H_{max}} n_h(H)\ln n_h(H) dH, \tag{1}$$

$$E = \int_0^{H_{max}} m_h g H n_h(H) dH + \int_0^{H_{max}} m_s g H n_s(H) dH. \tag{2}$$

The equilibrium NDFs can be determined by minimizing the free energy,

$$\delta F = \delta E - \delta(T_{seg} S) = 0, \tag{3}$$

subject to the constraints of particle number conservation at each height, i.e., $\delta n_s(H) = -\delta n_h(H)$, owing to the regular shape of the shear box. The solution of Eq. (3) yields a Boltzmann-like expression for the ratio between $n_s(H)$ and $n_h(H)$:

$$\frac{n_h(H)}{n_s(H)} = e^{\frac{\beta}{2}\Delta m g H} \tag{4}$$

where $\beta = 1/T_{seg}$ and $\Delta m = m_s - m_h$ is the mass difference between SPs and HPs.

To measure $T_{seg}$, we plot the logarithm of the left-hand side of Eq. (4) as a function of $H$ for all cases [insets of Figs. 2(c) and 2(d)]. In confined systems where pressure is uniform, $\ln[n_h(H)/n_s(H)]$ shows a linear dependence on $H$ within the height range $H/d \in [6,13]$, allowing us to extract $T_{\text{seg}}$ as the slope should equal $\frac{\Delta mg}{2T_{\text{seg}}}$. However, no such linear relationship can be recognized between $\ln[n_{\text{h}}(H)/n_{\text{s}}(H)]$ and $H$ for free surface systems. To resolve this discrepancy, we note that for free surface systems, the internal pressure exhibits hydrostatic behavior, varying linearly with depth, resulting in a nonuniform vertical pressure distribution. Instead, we plot $p \cdot \ln[n_{\text{h}}(H)/n_{\text{s}}(H)]$ as a function of $H$ and identified linear regimes for $H/d \in [13,17]$ [Figs. 2(c) and 2(d)]. The reappearance of Boltzmann-like behavior implies that the steady-state distribution of segregation is governed by $T_{\text{seg}}/p$ rather than $T_{\text{seg}}$, indicating different heights of the system equilibrate at identical $T_{\text{seg}}/p$. This issue is absent in confined systems where $p$ is constant. Similarly, we can once again extract $T_{\text{seg}}$ since the slope now equals $\frac{\Delta mg}{2(T_{\text{seg}}/p)}$.

As $T_{\text{seg}}/p$ has the dimension of volume, it is natural to examine its relationship with the temperature-like quantity in Edwards' volume ensemble, i.e., the compactivity $\chi$, which also has the dimension of volume. Following the standard fluctuation theorem, which relates energy fluctuations with heat capacity, we obtain $\chi$ similar to previous studies [28,30,31]. Across different pressure conditions and $\Gamma$, we observe a clear linear relationship between $\chi$ and $T_{\text{seg}}/p$, with a proportionality constant $\chi p/T_{\text{seg}} = 0.9$, as shown in Fig. 3. Considering experimental uncertainties, we can conclude that $T_{\text{seg}}/p$ equals $\chi$ in our systems. This finding demonstrates that although only the mixing entropy is considered in the free energy, the density-driven segregation process is still governed by the Edwards compactivity, originating from the configurational entropy. It is interesting to note that for free surface systems, different heights of the system equilibrate at identical $T_{\text{seg}}/p$ or $\chi$. This serves as, in fact, another direct confirmation of the zeroth law of the Edwards ensemble [32]. Consequently, the correct form of free energy should be $F = E/p + V - \chi S$ instead of the commonly known one $F = E + pV - T_{edw}S$, even though they may appear identical after normalization by a simple scaling factor $p$.

To understand why the Boltzmann regime is observed only within limited height ranges for both free surface and confined systems, we examine the self-intermediate scattering function $F_S(q, \delta\gamma)$ for particles at different heights [insets of Figs. 2(e) and 2(f)], where $q = 3.5$ denotes the wave vector corresponding to the first peak of the static structure factor. As shown in the insets of Figs. 2(e) and 2(f), $F_S(q, \delta\gamma)$ for both free surface and confined systems can be well-fitted by a Kohlrausch-Williams-Watts (KWW) function, $F_S(q, \delta\gamma) = \exp\left[-(\delta\gamma/\Delta\gamma_\alpha)^\lambda\right]$. We find that the corresponding structural relaxation time $\Delta\gamma_\alpha$, which measures how long a particle needs to change its neighborhood, exponentially increases with depth under free surface conditions, while $\Delta\gamma_\alpha$ remains almost constant in the bulk region of the system under confined conditions. Therefore, it is reasonable to suggest that when $\Delta\gamma_\alpha$ exceeds $\tau$, segregation processes at the corresponding heights fail to achieve steady states within the experimental time scale. Consequently, the Boltzmann-like distribution is realized only within finite height regions where $\Delta\gamma_\alpha < \tau$, which is consistent with our experimental observations [highlighted in blue region in Fig. 2].

In summary, we use X-ray tomography to study quasi-statically sheared granular systems with particles of different densities. We find that the steady-state particle number distributions, under various cyclic shear amplitudes and system pressures, can be well comprehended using Edwards statistics-based thermodynamic framework incorporating concepts like Edwards compactivity. Notably, the free energy now takes a new form $F = E/p + V - \chi S$ instead of the commonly known one $F = E + pV - TS$. We believe these results are important not only for deepening our understanding of segregation of granular materials, but also for laying the groundwork for the development of a thermodynamic framework for granular materials. It is reasonable to anticipate that additional terms, *e.g.*, elastic energy, and chemical potential, can be all included in this new framework. This expansion will facilitate future advancements in the development of constitutive models for granular materials.

The work is supported by the National Natural Science Foundation of China (No. 12274292).

Corresponding author

*yujiewang@sjtu.edu.cn

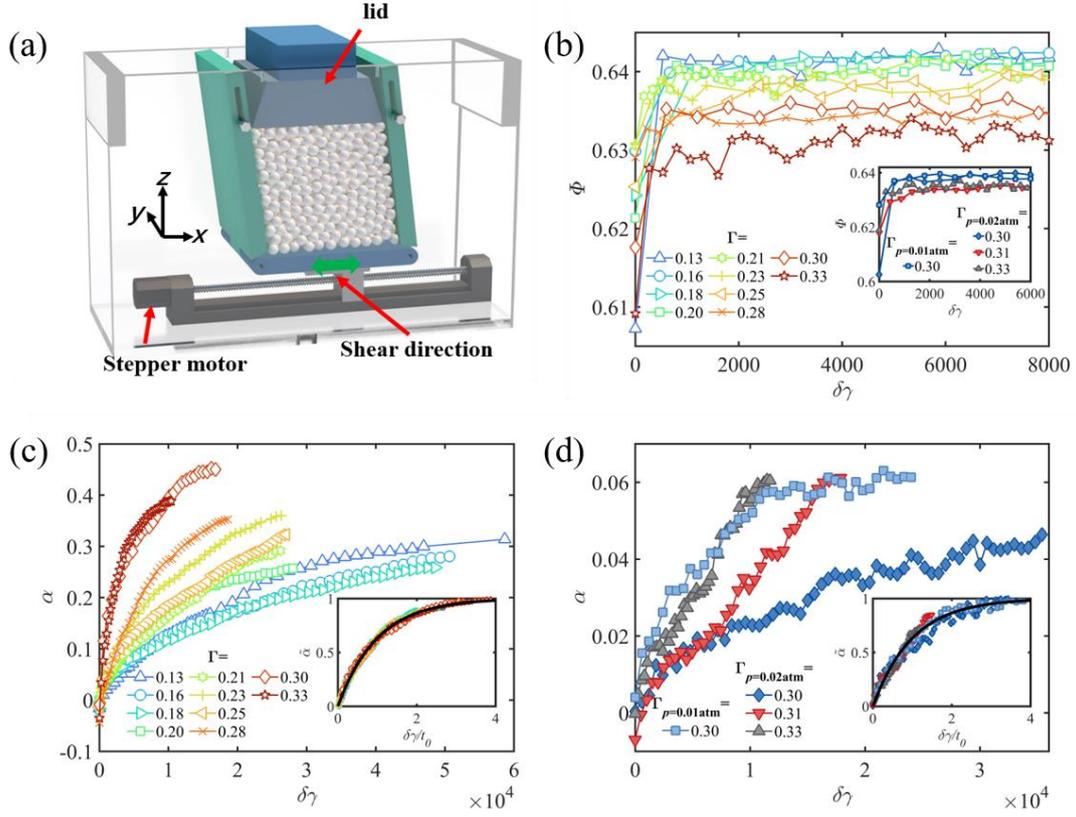

FIG. 1. (a) Schematic of the experimental setup. (b) Volume fraction $\phi$ as a function of $\delta\gamma$ for free surface systems with different $\Gamma$ (different symbols). The inset shows the same parametric plot for confined systems. (c, d) Degree of segregation $\alpha$ as a function of $\delta\gamma$ for free surface (c) and confined systems (d) with different $\Gamma$. Insets in (c) and (d) show the relation between $\alpha$ and $\delta\gamma/\tau$ (see main text) with an empirical fit $\alpha = 1 - \exp(-\delta\gamma/\tau)$ (solid curves).

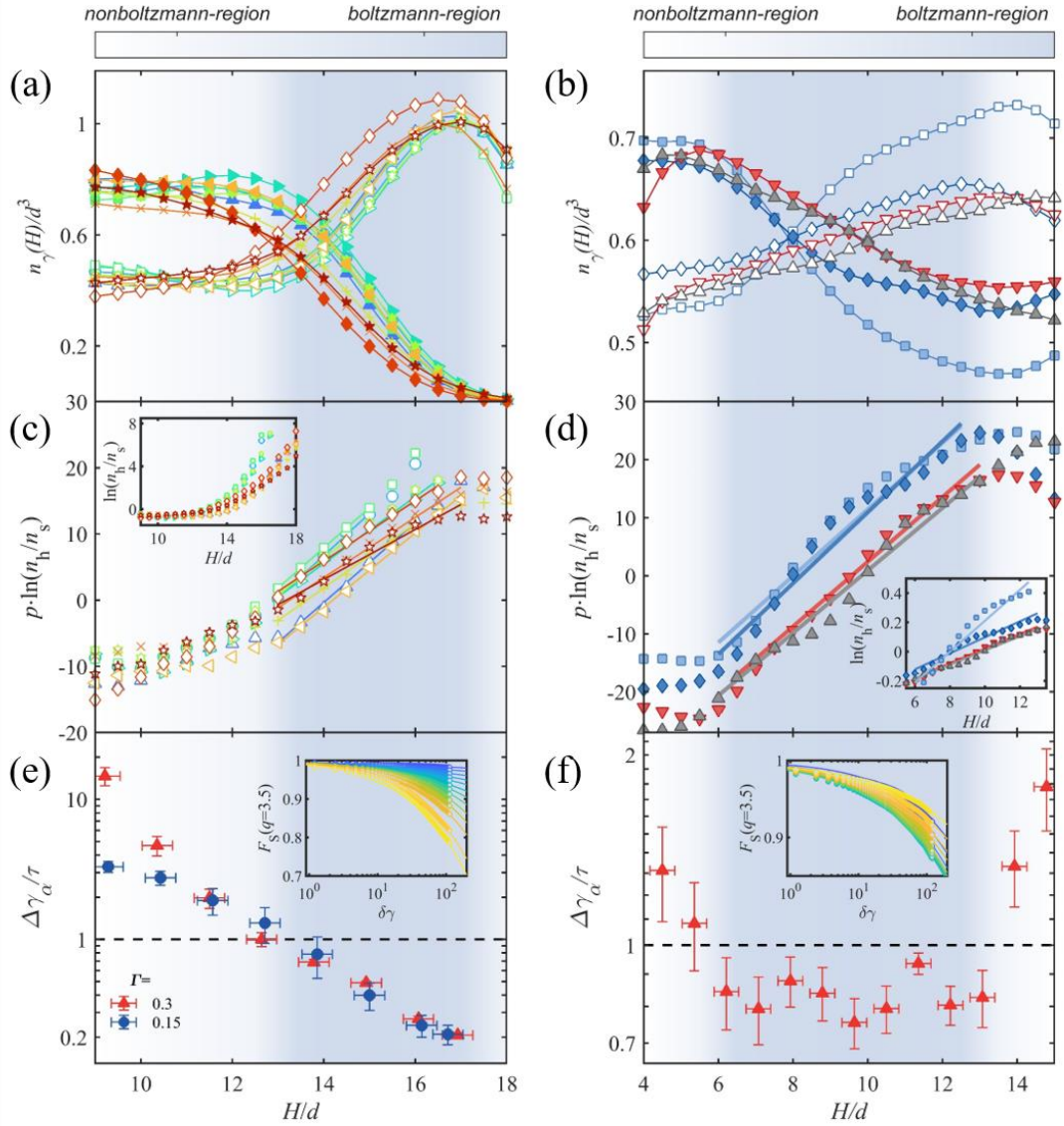

FIG. 2. (a, b) NDFs of HPs (open symbols) and SPs (filled symbols) at steady state as a function of height $H$ for free surface (a) and confined (b) systems. Different symbols represent different $\Gamma$. (c, d) Relationship between $p \cdot \ln(n_h / n_s)$, in unit of $\Delta mgd / 2V$, and $H/d$ for free surface (c) and confined (d) systems. The solid lines are the linear fit to the data in the Boltzmann regime. Insets in (c) and (d) show the relationship between $\ln(n_h / n_s)$ and $H/d$ for free surface and confined system. (e, f) Structural relaxation time $\Delta \gamma_\alpha$ divided by the characteristic time of segregation $\tau$ (see main text) as a function of $H$ for free surface systems (e) with $\Gamma = 0.15$

(blue) and 0.3 (red), and confined systems (f) with $\Gamma=0.3$ and $p$=0.02 atm. The light blue shaded region denotes the Boltzmann regime where $\Delta\gamma_\alpha$ is below $\tau$. Insets in (e) and (f) show the self-intermediate scattering function $F_S(q=3.5)$ as a function of $\delta\gamma$ for particles in different-height horizontal slices of free surface systems with $\Gamma=0.3$, and of confined systems with $\Gamma=0.3$ and $p$=0.02 atm. Slices with different heights (from top to bottom) are marked by different color symbols (from yellow to blue).

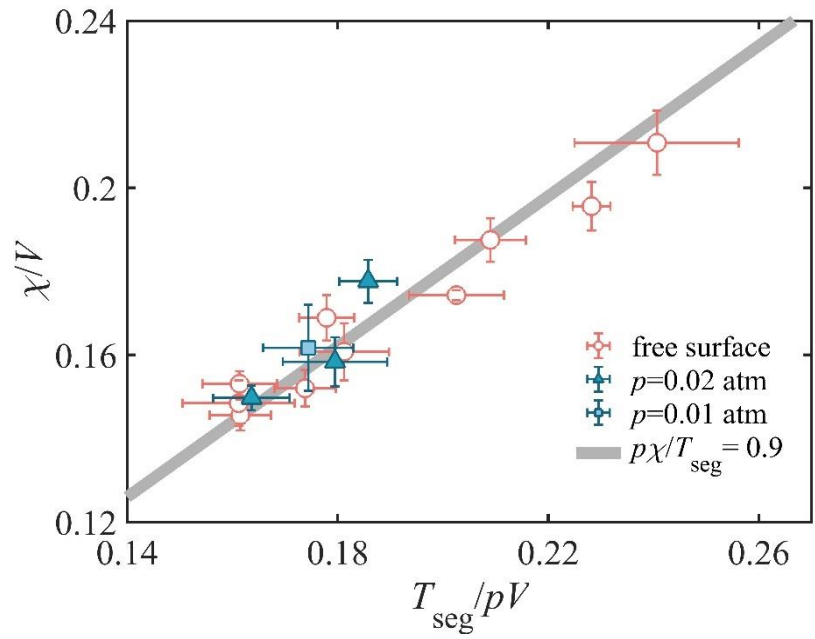

FIG. 3. Edwards' compactivity versus segregation temperature (in unit of particle volume $V$) for free surface (open symbols) and confined (filled symbols) systems with different $\Gamma$. The solid line represents the linear fit of $p\chi/T_{seg} = 0.9$.

# Supplemental Material for

# Edwards Thermodynamic Framework Controls Density Segregation in Cyclically Sheared Granular Materials

Haiyang Lu,[1] Houfei Yuan,[1] Shuyang Zhang,[1] Zhikun Zeng,[1] Yi Xing,[1] Jiazhao Xu,[1] Xin Wang,[1] and Yujie Wang[1,2,3,*]

[1]*School of Physics and Astronomy, Shanghai Jiao Tong University, Shanghai 200240, China*
[2]*State Key Laboratory of Geohazard Prevention and Geoenvironment Protection, Chengdu University of Technology, Chengdu 610059, China*
[3]*Department of Physics, College of Mathematics and Physics, Chengdu University of Technology, Chengdu 610059, China*

Corresponding author

[*]yujiewang@sjtu.edu.cn

**1. Edwards compactivity**

To obtain Edwards compactivity $\chi$, we prepare a series of packings with volume fractions $\phi$ ranging from RLP (random loose packing) to approximately RCP (random close packing). Firstly, we secure a reproducible initial steady state by driving the system through several thousand cycles of quasi-static cyclic shear with a shear amplitude of $\Gamma = 0.05$. Subsequently, we shear the system with uniform shear strain steps ($\Delta\Gamma = 0.17$) to a maximum strain of $\gamma = 0.63$ to obtain packings with different $\phi$. After each strain step, we obtain the three-dimensional (3D) packing structures by performing X-ray tomography. Figure S1(a) shows the evolution of $\phi$ as a function of shear strain $\gamma$, with $\phi$ reaching its steady-state value at $\gamma = 0.52$ [33]. By employing a standard hopper deposition protocol (HDP), we can also obtain packings in the RLP state with a corresponding volume fraction $\phi^r = 0.6011$, which aligns consistently with $\phi$ obtained from the direct shear [Fig. S1 (a)]. The inset of Fig. S1(a) depicts the steady-state volume fraction $\phi$ for systems with different $\Gamma$ under cyclic shear.

In Figure S1(b), we present volume fluctuations $\sigma_V^2$ as a function of $\phi$ under both direct shear and cyclic shear conditions. Data from both types of shear collapse onto a master curve. Subsequently, we fit these data with a quadratic polynomial as follows (solid line in Fig. S2):

$$\sigma_V^2 = 1.0576 - 3.1281\phi + 2.3377\phi^2. \tag{S1}$$

According to the Edwards volume ensemble, compactivity $\chi$ can be defined by the fluctuation method [30,31]:

$$\frac{1}{\chi(\phi)} - \frac{1}{\chi^r} = \int_{\phi^r}^{\phi} \frac{d\varphi}{\varphi^2 \sigma_V^2}, \tag{S2}$$

where $\phi^r$ is the packing fraction of the RLP state with an infinite $\chi^r$. Using the curve of $\sigma_V^2(\phi)$ obtained above, we can therefore determine $\chi$ of the packings.

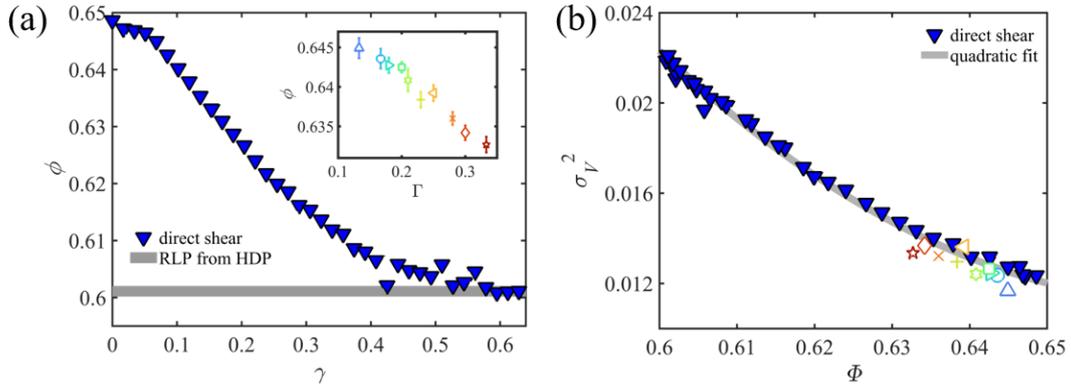

FIG. S1. (a) Volume fraction $\phi$ as a function of shear strain $\gamma$ under direct shear. The solid line marks the position of $\phi^r = 0.6011$ for RLP state obtained from the HDP. Inset: volume fraction $\phi$ as a function of strain amplitude $\Gamma$ at steady state under cyclic shear. (b) Volume fluctuation $\sigma_V^2$ as a function of $\phi$ in direct shear (filled symbols) and cyclic shear (open

symbols) experiments. The solid curve represents the quadratic polynomial fit of Eq. (S2).

## 2. Characteristic time of segregation

By fitting the relationship between the scaled degree of segregation $\alpha$ and the accumulated strain $\delta\gamma$ with $\alpha = 1 - \exp(-\delta\gamma/\tau)$, we obtain the characteristic time of segregation $\tau$ under different $\Gamma$, as illustrated in Fig. S2. It shows that $\tau$ monotonically decreases with increasing $\Gamma$ in free surface systems, while pressure exerts no significant influence.

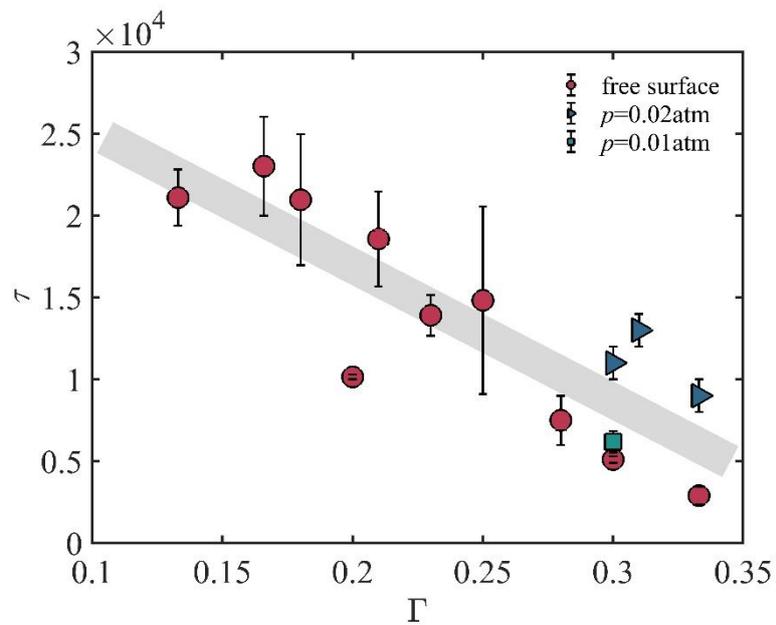

FIG. S2. Characteristic time of segregation $\tau$ as a function of $\Gamma$ for free surface system (circles), and confined system with $p$=0.02 atm (triangles) and $p$=0.01 atm (squares).